\pdfoutput=1

\documentclass[11pt]{article}

\usepackage[final]{acl}

\usepackage{times}
\usepackage{latexsym}
\usepackage{xspace}
\usepackage{amsmath}
\usepackage{booktabs}
\usepackage{graphics}
\usepackage{multirow}
\usepackage{pifont}
\usepackage{hyperref}
\usepackage{inconsolata}
\usepackage[hang,flushmargin]{footmisc}
\usepackage{bbm}

\title{Can't Hide Behind the API: Stealing Black-Box\\ Commercial Embedding Models}

\author{Manveer Singh Tamber, Jasper Xian, Jimmy Lin \\[1ex]
David R. Cheriton School of Computer Science,\\ University of Waterloo, Canada \\[1ex]
\texttt{\{mtamber, j5xian, jimmylin\}@uwaterloo.ca}}
\begin{document}
\maketitle

\begin{abstract}

Embedding models that generate dense vector representations of text are widely used and hold significant commercial value. 
Companies such as OpenAI and Cohere offer proprietary embedding models via paid APIs, but despite being ``hidden'' behind APIs, these models are not protected from theft.
We present, to our knowledge, the first effort to ``steal'' these models for retrieval by training thief models on text--embedding pairs obtained from the APIs.
Our experiments demonstrate that it is possible to replicate the retrieval effectiveness of commercial embedding models with a cost of under \$300. 
Notably, our methods allow for distilling from multiple teachers into a single robust student model, and for distilling into presumably smaller models with fewer dimension vectors, yet competitive retrieval effectiveness.
Our findings raise important considerations for deploying commercial embedding models and suggest measures to mitigate the risk of model theft.

\end{abstract}

\section{Introduction}

Embedding models are widely used and hold significant commercial value. 
Companies such as OpenAI and Cohere have developed competing proprietary models, accessible only through paid APIs.
By limiting direct access, companies retain exclusive control over their models and monetize their usage.
While these APIs provide convenient access to users, they may also hold some disadvantages, such as potential additional costs and concerns about data privacy and security.

This raises an intriguing question:\ Can these embedding models be ``stolen'' using distillation techniques?

Stealing a model can offer several potential benefits to different actors. 
It can reduce costs, enhance data privacy and security by eliminating reliance on external APIs, enable fine-tuning for specific needs, and potentially lower query latency for downstream applications.
Additionally, competitors may attempt model theft to gain a strategic advantage.
Stealing retrieval models could also appeal to adversaries engaged in black-hat search engine optimization (SEO)~\cite{imitationattack}. After stealing an embedding model, an adversary could analyze the stolen model and use it to craft adversarial passages designed to artificially boost rankings, with the expectation that these manipulated passages also rank highly for the query under the original embedding model.

To our knowledge, this work is the first to explore stealing embedding models for retrieval. 
We query APIs from OpenAI and Cohere to collect text--embedding input--output pairs, which are used to train our thief models. 
We demonstrate that strong embedding models, initialized from BERT base and BERT large~\cite{devlin-etal-2019-bert}, can be trained to accurately and cost-effectively mimic their API-based counterparts. 
The stolen models demonstrate strong effectiveness across out-of-domain retrieval tasks and datasets, despite training exclusively on the MSMARCO v1 passage ranking dataset.
Additionally, we explore distilling from both OpenAI and Cohere models into a single student model, which shows promise for building strong embedding models that distill from multiple teachers simultaneously.

\section{Background}

\paragraph{Dense Retrieval}
Dense retrieval involves representing both queries and documents as dense vector embeddings in a high-dimensional vector space, aiming for query embeddings to be close to relevant document embeddings while pushing irrelevant ones farther away~\cite{conceptualframework, karpukhin-etal-2020-dense}.
Embedding models, such as those provided by OpenAI and Cohere, are tasked with generating embeddings for queries and documents across diverse topics and domains, including those beyond the models' original training data. The models can then be used for searching a pre-encoded corpus of documents with a query embedding by finding nearest neighbours using measures such as the dot product or cosine similarity. 

\paragraph{Model Distillation.}
Knowledge distillation work in machine learning follows from \citet{hinton2015distilling} where a teacher's predictions in classification tasks were used as soft targets to train a student model.

DistilBERT~\cite{sanh2019distilbert} demonstrated that a smaller model can be distilled from BERT while retaining most of the capabilities of the larger teacher model. 
This was achieved using a combination of a distillation loss to align soft target probabilities for classification, a masked language modeling loss, and a cosine embedding loss to align hidden state vectors.

\citet{Xie2023} proposed a simplified distillation approach for BERT sentence embedding models, where a smaller model with fewer transformer layers was initialized using the teacher’s weights.
An embedding loss was applied to align hidden state vectors and output embeddings.

Multilingual sentence embedding distillation has also been explored~\cite{reimers-gurevych-2020-making}, where student models were trained to match the output embeddings of a teacher model for English text and its translations. BERT models were used for the student and teacher encoders.

Other work has shown that BERT’s knowledge can be distilled into a single-layer BiLSTM, significantly reducing model parameters and inference time~\cite{distillnn}.

Unlike the works mentioned above, our setup operates in a black-box setting. We distill teacher models without knowing how they are initialized and trained, or the specifics of their training data.
We do not have access to the activations in the teacher's hidden layers.
Some existing black-box distillation work~\cite{blackboxkd, ma2024aligninglogitsgenerativelyprincipled} focuses on models trained on classification tasks where student models align their classification probability distributions with teacher models. 
In our approach, only a single embedding vector is returned by the APIs, so we train using a loss that aligns the final embeddings produced by the student models with those from the API models. 

\paragraph{Model-Stealing.}
Research has investigated model-stealing attacks that use input--output pairs obtained by querying APIs. 
These pairs are used to train equivalent or near-equivalent models, often with little knowledge of the training data or the architecture behind the black-box API~\cite{tramer2016stealing, cloudleak}.

In computer vision, recent work has successfully stolen image encoders by training models to replicate the embeddings generated by the target encoders~\cite{sha2023can, stolenencoder}, with the stolen encoders evaluated on downstream classification tasks.

In NLP, transformer-based models built with BERT have also been successfully stolen, with extracted models performing nearly as well as the original API models for tasks such as sentiment classification, question answering, and natural language inference~\cite{krishna2019thieves}.

\citet{dziedzic2023sentence} examined the theft of text embedding models, where the victim models were initialized using either TinyBERT, BERT base, or RoBERTa large and fine-tuned on natural language inference datasets. The stolen models were then evaluated on the SentEval benchmark~\cite{conneau2018senteval}.

However, there remains room to explore the theft of text embedding models behind APIs. 
First, the previous work evaluated their models on the simpler SentEval benchmark, which predates the BERT era in NLP. 
Moreover, their work concentrated on models fine-tuned for natural language inference, rather than embedding models for retrieval tasks.
Additionally, the work assumes full knowledge of the victim model’s architecture and training data, allowing them to use similar architectures and training datasets. This assumption does not hold in real-world API settings, where such details are often unknown.

A recent model-stealing attack was proposed that extracted the embedding projection layer of transformer models through targeted queries~\cite{carlini2024stealing}. 
The authors found that they could determine the hidden dimension size of black-box OpenAI models. 
Although the authors noted no immediate practical application for this attack, it is a novel and interesting approach.

\begin{table*}[ht!]
\centering
\resizebox{\linewidth}{!}{ \begin{tabular}{llcrrrr}
\toprule
     \multicolumn{1}{c}{\multirow{2}{*}{\textbf{Company}}}
     & \multicolumn{1}{c}{\multirow{2}{*}{\textbf{Model}}} & \multicolumn{1}{c}{\textbf{Input Type}} & \multicolumn{1}{c}{\textbf{Embedding}} & \multicolumn{1}{c}{\textbf{Max Tokens}} & \multicolumn{1}{c}{\textbf{Cost /}} & \multicolumn{1}{c}{\textbf{MTEB Retrieval}} \\

     & & \textbf{API Argument} & \multicolumn{1}{c}{\textbf{Dimensions}} & \multicolumn{1}{c}{\textbf{Length Limit}} & \multicolumn{1}{c}{\textbf{1M tokens}} & \multicolumn{1}{c}{\textbf{Average nDCG@10}} \\ 
\midrule
    OpenAI & text-embedding-3-large & \ding{55} & 3072 & 8192 & \$0.13 & 55.4 \\
    Cohere & embed-english-v3.0 & \ding{51} & 1024 & 512 & \$0.10 & 55.0  \\

\bottomrule
\end{tabular}}
\caption{Comparing OpenAI and Cohere's flagship embedding models.}
\label{tab:apis}
\end{table*}

\section{Methods}

In our work, we follow the real-world scenario of stealing a commercial embedding model. We assume the attacker has only black-box access to the model, meaning they are unaware of the model's exact initialization, architecture, or training process. The attacker may only query the model at a modest cost and make assumptions by leveraging publicly available information and tools from the company providing the API to replicate the model. As we later show, publicly available information can potentially be very valuable for the attacker.

\subsection{Victim Models}
\label{API}

In this work, we target two live victim embedding models behind APIs: OpenAI's text-embedding-3-large and Cohere's embed-english-v3.0, the flagship embedding models of these companies. Key differences between the models are summarized in Table~\ref{tab:apis}.

Cohere’s model distinguishes between queries and documents based on the \texttt{input\_type} argument provided to the API, while OpenAI’s model makes no such distinction. 
Cohere’s API offers four options for \texttt{input\_type}: \texttt{search\_document}, \texttt{search\_query}, \texttt{classification}, and \texttt{clustering}. 
Although the exact differences in embedding queries and documents are unclear, we assume the model is prompted differently depending on the chosen \texttt{input\_type}.

Both models produce embeddings with different dimensionality. 
Cohere’s model outputs 1024-dimensional embeddings, while OpenAI’s model generates embeddings with 3072 dimensions. 
OpenAI's model, however, uses Matryoshka Representation Learning~\cite{mrl}, allowing embeddings to be shortened by removing numbers from the end, trading off some effectiveness for reduced vector dimensions. 
Additionally, OpenAI’s API allows for a maximum input length of 8192 tokens, whereas Cohere’s API limits input length to 512 tokens.

We suspect Cohere’s model is based on BERT, as they provide a publicly available tokenizer on HuggingFace\footnote{\url{https://huggingface.co/Cohere/Cohere-embed-english-v3.0}} that uses the \texttt{BertTokenizer} class.
The model's 512-token limit further supports this, and the 1024-dimensional output suggests a variant of BERT large.
It is more difficult to guess which model OpenAI uses to initialize their embedding model, but it is likely not based on BERT given the larger token limit and embedding dimensions.
Decoder-only LLMs have been used as embedding models that handle long sequences and produce embeddings of a large number of dimensions~\cite{ma2023fine}.
The T5 model’s encoder has also been effective in generating embeddings~\cite{ni2021large}. That said, some research has demonstrated success in adapting BERT to handle longer sequences~\cite{nussbaum2024nomic}.

In terms of cost, OpenAI’s model is slightly more expensive at \$0.13 per million tokens compared to Cohere’s \$0.10 per million tokens, though their tokenizers differ. Both companies report scores on the retrieval task of the Massive Text Embedding Benchmark (MTEB) \cite{muennighoff-etal-2023-mteb}, with OpenAI’s model performing slightly better than Cohere’s, but still achieving comparable results as shown in Table~\ref{tab:apis}. Interestingly, research by \citet{merrick2024arctic} has shown that embedding models trained from BERT base can achieve competitive performance, with an MTEB Retrieval Average nDCG@10 of $55.1$, comparable to these commercial models. This motivates the distillation of these commercial embedding models into BERT models.

\subsection{Model Architecture} \label{Architecture}

Our embedding models are initialized with the uncased variants of BERT base (110M parameters) and BERT large (340M parameters), chosen for their ease of training and use. 
An interesting aspect of using BERT is that, since Cohere’s model is likely initialized from a BERT variant and OpenAI’s model is likely not, this allows us to study distillation from both a (presumed) BERT and non-BERT teacher into a BERT student.

We use the same model for both queries and documents, which has been shown to work well in dense retrieval tasks~\cite{izacard2021contriever, xiong2020approximate, reimers-gurevych-2019-sentence}.

As explained in Section~\ref{API}, Cohere’s API differentiates between queries and documents using the \texttt{input\_type} argument.
To distill from both OpenAI’s and Cohere’s model, we prepend either ``Query: '' or ``Document: '' to the text based on whether it represents a query or a document.

A challenge with distillation is the dimensionality mismatch: BERT base produces 768-dimensional embeddings, BERT large produces 1024-dimensional embeddings, while OpenAI’s model outputs 3072-dimensional embeddings, and Cohere’s model outputs 1024-dimensional embeddings. 
To address this, we extract embeddings by averaging the hidden representations of the last layer, then apply a learnable linear transformation to match the dimensions of the target embeddings.

Interestingly, because the linear transformation is applied after generating the intermediate representations, the original 768-dimensional embeddings from BERT base and the 1024-dimensional embeddings from BERT large can be used directly.
As demonstrated in Section~\ref{BottleneckRepresentations}, these embeddings offer comparable retrieval effectiveness to the longer (or sometimes equal length) transformed embeddings. 
For efficiency, we prioritize using the shorter embeddings for retrieval whenever possible.

\subsection{Training Data}
Ideally, we would train our thief models with as much data as possible. 
Given the reasonable costs of using commercial embedding APIs, it is feasible to encode large amounts of text, enabling the generation of substantial training data for distillation.
For practical reasons such as storage and costs, we impose a sensible limit on the amount of training data used.

Since models may embed queries and passages differently, it is necessary to train with both.
We train our thief models to replicate the embeddings produced by the API models for queries and passages from the MSMARCO v1 passage ranking dataset~\cite{bajaj2016ms}, a widely used dataset in information retrieval. 
The dataset’s 8.8 million passages and 809k queries cover a diverse range of topics, making it ideal for training models to generalize across various domains.
Passages from the MSMARCO v1 passage corpus are extracted from web documents retrieved by Bing and should capture a diversity of content, grammar, and style in text.
To enhance data efficiency, we remove duplicate passages where some are prefixes or suffixes of others, reducing the total number of passages to approximately 8.4 million. We reserve 400k passages and 100k queries as a development set, selected from the end of the passage corpus and query set based on their IDs. We train our thief models with either 100k, 400k, or the full set of 8.7 million text–embedding pairs to test varying the amount of training data.
Each subset is randomly sampled to ensure it represents the broader dataset and includes both queries and passages.

We estimate the cost of generating these query and passage embeddings through the APIs to be approximately \$88 using OpenAI’s text-embedding-3-large model and \$68 using Cohere’s embed-english-v3.0 model.

\subsection{Cosine Distance Loss}
A simple yet effective way to distill embeddings from a teacher to a student is using an average cosine distance loss calculated using the term:
$$-\frac{1}{n}\sum_{i=1}^n\frac{t_i \cdot s_i}{\Vert t_i\Vert_2 \cdot \Vert s_i \Vert_2}$$ where $t_i$ is the embedding produced by the API model (teacher) for some text, either a query or a document, and $s_i$ is the embedding produced by our thief model (student) for the same text. 
Training with this loss aligns the direction of the two vectors. 
Since the APIs provide normalized vector embeddings and the embeddings from our models are normalized as well, training with this loss function ensures that the vectors closely match.

\subsection{Contrastive Loss}

Instead of the cosine distance loss, \citet{sha2023can} proposed using a contrastive loss to steal image encoders:
$$-\frac{1}{n}\sum_{i=1}^n \log{\frac{e ^ {\frac{sim(t_i, s_i)}{\tau}}}{\sum_{j=1}^n e ^ {\frac{sim(t_j, s_i)}{\tau}} + \sum_{j=1, j\neq i}^n e ^ {\frac{sim(s_j, s_i)}{\tau}}}}$$

\noindent where $sim()$ is the cosine similarity of the two vectors.
Their work argued that contrastive signals in the loss helped improve stealing effectiveness over a simpler MSE loss. We compare training with the simpler cosine distance loss and this loss in Section~\ref{ContrastiveLoss}
 \label{ContrastiveLossFunction}

\section{Experimental Setup}
\begin{table*}[ht!]
\centering
\resizebox{1.0\linewidth}{!}{ 
\begin{tabular}{l|cc|cc|cc|cc|cc|cc|cc}
\toprule
\toprule
 & \multicolumn{2}{c|}{\textbf{API}} & \multicolumn{6}{c|}{\textbf{BERT Base Thief}} & \multicolumn{6}{c}{\textbf{BERT Large Thief}}  \\
 & \multicolumn{2}{c|}{\textbf{}} & \multicolumn{2}{c|}{\textbf{(Q Only)}} & \multicolumn{2}{c|}{\textbf{(P Only)}} & \multicolumn{2}{c|}{\textbf{(Q\&P)}} & \multicolumn{2}{c|}{\textbf{(Q Only)}} & \multicolumn{2}{c|}{\textbf{(P Only)}} & \multicolumn{2}{c}{\textbf{(Q\&P)}} \\
  & nDCG & Recall & nDCG & Recall  & nDCG & Recall & nDCG & Recall & nDCG & Recall  & nDCG & Recall & nDCG & Recall \\

\midrule
\underline{\textbf{Cohere}} \\
DL19                   & 69.6 & 64.8 & 69.2 & 63.3 & 70.8 & 64.2 & 70.6 & 64.4 & 69.9 & 64.2 & 70.8 & 65.1 & 70.6 & 64.6 \\
DL20                   & 72.5 & 72.8 & 71.9 & 72.0 & 72.2 & 72.1 & 72.2 & 72.0 & 72.5 & 72.0 & 72.3 & 72.9 & 72.9 & 72.7 \\
TREC-COVID             & 81.8 & 15.9 & 79.2 & 15.2 & 70.6 & 13.0 & 70.6 & 12.3 & 81.5 & 15.9 & 74.5 & 14.1 & 73.5 & 14.0 \\
BioASQ                 & 45.7 & 67.9 & 44.0 & 66.2 & 32.2 & 56.2 & 32.3 & 55.3 & 44.1 & 67.7 & 34.9 & 59.4 & 34.9 & 59.0 \\
NFCorpus               & 38.6 & 35.1 & 38.6 & 35.0 & 35.3 & 33.2 & 35.1 & 33.3 & 38.4 & 34.9 & 36.8 & 33.6 & 36.6 & 34.2 \\
NQ                     & 61.6 & 95.6 & 61.3 & 95.4 & 50.0 & 87.7 & 49.5 & 87.4 & 61.9 & 95.6 & 50.6 & 87.9 & 50.7 & 88.0 \\
HotpotQA               & 70.7 & 82.3 & 68.7 & 80.6 & 67.3 & 80.1 & 66.2 & 78.9 & 69.7 & 81.5 & 67.6 & 80.6 & 67.3 & 79.9 \\
FiQA                   & 42.1 & 73.6 & 41.4 & 72.5 & 36.8 & 69.8 & 36.6 & 69.3 & 41.9 & 73.4 & 37.1 & 71.1 & 36.9 & 70.8 \\
Signal-1M              & 26.3 & 28.3 & 24.7 & 26.9 & 24.5 & 29.1 & 24.4 & 28.0 & 25.0 & 28.2 & 24.1 & 28.4 & 24.3 & 28.5 \\
TREC-NEWS              & 50.4 & 54.3 & 49.8 & 53.5 & 47.3 & 50.2 & 47.0 & 49.0 & 49.9 & 54.1 & 50.1 & 52.0 & 49.9 & 51.1 \\
Robust04               & 54.1 & 41.7 & 53.5 & 41.5 & 52.3 & 39.9 & 52.1 & 40.0 & 53.7 & 41.4 & 54.5 & 41.6 & 53.9 & 41.2 \\
Arguana                & 54.0 & 98.2 & 43.3 & 98.9 & 39.6 & 96.9 & 37.8 & 96.4 & 43.1 & 98.9 & 41.1 & 98.0 & 38.9 & 97.1 \\
Touché-2020            & 32.6 & 51.6 & 33.3 & 51.2 & 29.8 & 47.0 & 30.1 & 46.8 & 32.8 & 51.9 & 32.2 & 48.7 & 31.8 & 48.5 \\
Quora                  & 88.7 & 99.6 & 88.0 & 99.6 & 85.2 & 99.4 & 85.3 & 99.4 & 88.3 & 99.6 & 83.8 & 99.3 & 83.9 & 99.3 \\
DBPedia                & 43.4 & 53.6 & 43.0 & 52.6 & 42.1 & 52.8 & 41.4 & 52.6 & 43.4 & 53.4 & 41.4 & 53.5 & 41.7 & 53.3 \\
SCIDOCS                & 20.3 & 45.1 & 19.4 & 42.6 & 17.7 & 41.5 & 16.7 & 39.6 & 19.6 & 43.0 & 18.2 & 43.1 & 17.5 & 41.6 \\
FEVER                  & 89.0 & 96.5 & 86.7 & 96.3 & 70.6 & 94.2 & 68.7 & 93.8 & 87.8 & 96.4 & 71.7 & 94.7 & 70.7 & 94.3 \\
Climate-FEVER          & 25.9 & 58.1 & 30.2 & 63.8 & 15.7 & 44.7 & 18.8 & 51.4 & 29.1 & 62.6 & 15.9 & 44.8 & 18.5 & 50.7 \\
SciFact                & 71.8 & 96.3 & 70.3 & 94.9 & 68.0 & 94.0 & 67.1 & 93.9 & 71.6 & 96.3 & 69.9 & 95.7 & 69.3 & 95.6 \\
\midrule
\textbf{TREC-DL Average}    & 71.1 & 68.8 & 70.6 & 67.7 & 71.5 & 68.2 & 71.4 & 68.2 & 71.2 & 68.1 & 71.6 & \underline{\textbf{69.0}} & \underline{\textbf{71.8}} & 68.7 \\
\textbf{BEIR Average}  & \textbf{52.8} & 64.3 & 51.5 & 63.9 & 46.2 & 60.6 & 45.9 & 60.4 & \underline{51.9} & \underline{\textbf{64.4}} & 47.3 & 61.6 & 47.1 & 61.6 \\
\midrule
\midrule
\underline{\textbf{OpenAI}} \\
DL19          & 71.7 & 64.7 & 67.7 & 62.8 & 70.4 & 64.6 & 68.3 & 63.6 & 68.2 & 64.6 & 69.3 & 64.8 & 68.3 & 64.3 \\
DL20          & 71.6 & 76.1 & 69.5 & 73.7 & 70.6 & 75.5 & 68.9 & 73.8 & 70.0 & 73.2 & 70.4 & 75.5 & 69.3 & 73.3 \\
TREC-COVID    & 76.9 & 15.6 & 51.9 & 10.9 & 73.3 & 13.7 & 64.9 & 11.8 & 59.9 & 11.2 & 76.5 & 14.8 & 68.7 & 12.6 \\
BioASQ        & 41.0 & 66.2 & 25.9 & 48.2 & 30.2 & 55.0 & 25.2 & 45.2 & 28.1 & 50.2 & 33.3 & 57.9 & 28.4 & 48.8 \\
NFCorpus      & 41.8 & 39.9 & 40.1 & 38.7 & 38.5 & 36.5 & 36.9 & 35.2 & 40.6 & 38.8 & 40.2 & 37.3 & 38.9 & 35.9 \\
NQ            & 58.6 & 95.7 & 52.2 & 91.8 & 50.7 & 92.0 & 45.0 & 87.3 & 54.7 & 93.1 & 54.0 & 93.3 & 49.7 & 89.9 \\
HotpotQA      & 69.7 & 85.8 & 49.7 & 71.2 & 60.7 & 80.7 & 49.5 & 70.4 & 53.8 & 74.6 & 63.6 & 82.2 & 54.6 & 74.0 \\
FiQA          & 54.7 & 84.3 & 47.5 & 79.3 & 39.5 & 75.5 & 36.1 & 70.5 & 49.0 & 80.8 & 43.7 & 77.9 & 40.0 & 74.6 \\
Signal-1M     & 27.1 & 32.2 & 24.9 & 28.8 & 26.2 & 31.2 & 26.4 & 28.6 & 25.2 & 29.9 & 26.3 & 30.9 & 27.3 & 29.3 \\
TREC-NEWS     & 52.2 & 58.1 & 47.7 & 55.1 & 44.7 & 49.6 & 40.7 & 47.3 & 49.2 & 55.2 & 49.7 & 51.9 & 47.3 & 51.3 \\
Robust04      & 58.0 & 46.0 & 52.9 & 43.2 & 54.7 & 41.5 & 51.0 & 39.5 & 54.8 & 43.7 & 57.0 & 43.8 & 54.3 & 42.4 \\
Arguana       & 40.3 & 99.1 & 41.7 & 98.7 & 39.6 & 98.1 & 38.4 & 97.3 & 42.0 & 98.5 & 40.9 & 98.7 & 40.1 & 98.0 \\
Touché-2020   & 29.2 & 49.2 & 29.8 & 51.2 & 29.1 & 49.4 & 26.5 & 48.3 & 30.0 & 50.9 & 30.4 & 50.6 & 29.9 & 50.6 \\
Quora         & 88.9 & 99.6 & 86.8 & 99.6 & 87.1 & 99.4 & 87.5 & 99.4 & 87.9 & 99.6 & 87.9 & 99.5 & 88.0 & 99.5 \\
DBPedia       & 43.4 & 56.6 & 39.3 & 52.6 & 40.4 & 52.4 & 38.3 & 51.1 & 40.1 & 53.7 & 41.7 & 53.9 & 39.1 & 53.1 \\
SCIDOCS       & 22.8 & 50.0 & 17.5 & 42.1 & 17.7 & 43.3 & 15.3 & 38.5 & 18.6 & 43.4 & 19.2 & 45.4 & 17.3 & 41.1 \\
FEVER         & 84.7 & 96.6 & 72.2 & 92.1 & 76.6 & 95.4 & 66.0 & 91.2 & 75.3 & 93.3 & 81.6 & 96.0 & 74.9 & 94.1 \\
Climate-FEVER & 27.2 & 62.8 & 27.7 & 61.2 & 26.1 & 60.3 & 26.6 & 60.1 & 28.1 & 62.1 & 26.9 & 61.1 & 27.9 & 60.4 \\
SciFact       & 76.1 & 97.7 & 67.9 & 94.7 & 71.3 & 97.0 & 65.7 & 94.9 & 68.9 & 95.3 & 72.6 & 96.0 & 67.5 & 95.3 \\
\midrule
\textbf{TREC-DL Average}    & \textbf{71.7} & \textbf{70.4} & 68.6 & 68.3 & \underline{70.5} & 70.1 & 68.6 & 68.7 & 69.1 & 68.9 & 69.9 & \underline{70.2} & 68.8 & 68.8 \\
\textbf{BEIR Average}  & \textbf{52.5} & \textbf{66.8} & 45.6 & 62.3 & 47.4 & 63.0 & 43.5 & 59.8 & 47.4 & 63.2 & \underline{49.7} & \underline{64.2} & 46.7 & 61.8 \\
\bottomrule
\bottomrule
\end{tabular}
}
\caption{nDCG@10 and Recall@100 scores on TREC-DL and BEIR datasets, comparing Cohere and OpenAI API models with BERT base and BERT large thief models across three retrieval settings: {(Q Only)} (thief encodes queries), {(P Only)} (thief encodes passages), and {(Q\&P)} (thief encodes both). Best average scores are in bold, with top scores for thief settings underlined.
}
\label{tab:full}
\end{table*}

\subsection{Model Training Details}
Model training and evaluation were performed on single 48GB RTX 6000 Ada GPUs using the \texttt{bfloat16} data type for faster-mixed precision evaluation and training.
To maximize GPU utilization while staying within the 48GB VRAM limit, all models used a batch size of 256.
A dropout rate of 10\% was used as it was found to lead to better effectiveness, particularly in data-limited settings. 
Training was performed using the AdamW optimizer with a default weight decay of 0.01 and a linear warmup over 50 steps, while a learning rate of 4e-5 was chosen to maintain stability, as higher values led to instability.

We observe that the models continue to improve over many training epochs. For datasets limited to 400k samples or fewer, models were trained for up to 200 epochs. 
For the full dataset of approximately 8.7 million samples, BERT base models were trained for 50 epochs, while BERT large models were limited to 40 epochs. In all cases, the best-performing model was selected based on loss on the dev set.
While continued training might yield further improvements, we limited training epochs in the interest of time.

When distilling from either victim model, training with a single 48GB RTX 6000 Ada GPU took approximately 6 hours for 100k samples and 20 hours for 400k samples. 
For the full dataset of 8.7 million samples, training BERT base required around 104 hours, while BERT large training took approximately 268 hours.

\begin{table*}[t!]
\centering
\resizebox{0.65\linewidth}{!}{
\begin{tabular}{l|c|ccc|c|ccc}
\toprule
\toprule
 & \multicolumn{4}{c|}{\textbf{Cohere}} & \multicolumn{4}{c}{\textbf{OpenAI}} \\
 
  & \multicolumn{1}{c|}{\textbf{API}} & \multicolumn{3}{c|}{\textbf{Thief}} & \multicolumn{1}{c|}{\textbf{API}} & \multicolumn{3}{c}{\textbf{Thief}} \\
  &  & \textbf{100k} & \textbf{400k} & \textbf{8.7M} &  & \textbf{100k} & \textbf{400k} & \textbf{8.7M} \\

\midrule
DL19        & 69.6 & 67.2 & 68.9 & 70.6 & 71.7 & 58.2 & 63.7 & 68.3 \\
DL20        & 72.5 & 67.7 & 70.7 & 72.2 & 71.6 & 60.6 & 66.1 & 68.9 \\
\midrule
\textbf{TREC-DL Average} & 71.1 & 67.5 & 69.8 & \textbf{71.4} & \textbf{71.7} & 59.4 & 64.9 & 68.6 \\
\bottomrule
\bottomrule
\end{tabular}
}
\caption{nDCG@10 scores on TREC-DL datasets with varying distillation training samples, testing retrieval by encoding both queries and passages.
}
\label{tab:starve}
\end{table*}

\subsection{Evaluation}

Since the models were trained on MSMARCO data, we evaluate in-domain retrieval using the TREC 2019~\cite{craswell2020overview} and 2020~\cite{craswell2021overview} Deep Learning (TREC-DL) Tracks from the MSMARCO v1 passage ranking task, and we evaluate out-of-domain retrieval using the BEIR benchmark~\cite{thakur2021beir}, which contains a variety of retrieval topics, query types, and sources of passage corpora.

Given that the thief encoders are intended to approximate the victim models that they steal from, we assess their effectiveness across three distinct settings. 
In the \textbf{(Q only)} setting, our models encode only the queries, with retrieval performed on corpora encoded by the original API embedding model. 
The \textbf{(P only)} setting involves our models encoding only the passages, while retrieval is conducted using queries encoded by the API model. 
The \textbf{(Q \& P)} setting has our models encode both queries and passages for retrieval.
While the (Q \& P) setting measures the retrieval effectiveness of the stolen model, the (Q only) setting also has practical relevance, as it allows for querying a corpus already encoded by the original model with reduced latency, potential cost savings, and greater data privacy and security for queries.

For retrieval effectiveness, we report mean nDCG@10 (Normalized Discounted Cumulative Gain at rank cutoff 10) and R@100 (Recall at rank cutoff 100) over queries for each retrieval task.

\section{Results}

\subsection{Training with All Available Data}

Table~\ref{tab:full} compares the retrieval effectiveness of our thief models trained on all available training samples. These models demonstrate strong effectiveness across MSMARCO and BEIR benchmarks, generalizing well to out-of-domain tasks despite training exclusively on MSMARCO v1 passage ranking data.

As expected, the BERT large thief models achieve stronger effectiveness scores than the BERT base thief models. While taking longer to train, the BERT large models are better able to approximate the commercial embedding models.

In the TREC-DL in-domain test collections, the score disparities between the thief settings and the API are generally smaller than in the BEIR collections. In some cases, the TREC-DL scores for Cohere’s thief settings even exceed those of the API. While BEIR scores remain strong, the thief models typically fall short of the API models. 
This is unsurprising because learning to target zero-shot retrieval is a challenging problem. Dense models fine-tuned using MSMARCO have previously been shown to perform strongly in-domain on MSMARCO, but underperform BM25 baselines out-of-domain on BEIR~\cite{thakur2021beir}. 
However, impressively, the thieves consistently achieve nDCG@10 and R@100 scores close to those of the API across diverse BEIR tasks, indicating effective generalization.
This is an interesting result, as it suggests that the distilled models can attain strong effectiveness across diverse retrieval tasks.

Comparing the Cohere and OpenAI thief settings, although inconsistently, the Cohere thief settings often lead in nDCG@10 scores, whereas OpenAI thief settings often have stronger recall@100 scores. 
While the OpenAI thieves' higher recall scores may be due to the higher recall scores from the OpenAI embedding model, we further examine how stealing from Cohere's and OpenAI's models compares in Section~\ref{comparisoncohereopenai}.

On BEIR tasks, encoding only queries or only passages with the thief models generally yields better retrieval effectiveness than encoding both with the thieves. This likely reflects the thief models’ imperfect approximation of the API models.
The API models are expected to be better retrieval models than the thieves because they are fine-tuned for retrieval. By training our thieves to mimic these API models, they approach the retrieval effectiveness of the API models.
Encoding only queries or passages with the thief models reduces the potential for compounding errors, resulting in stronger effectiveness. Notably, aside from the BERT large thief in the (Q only) setting achieving a higher average recall score than the full Cohere API setting, the API models maintain higher average retrieval scores.

\subsection{Training with Limited Data}

Table~\ref{tab:starve} shows the retrieval effectiveness of the thief models when the amount of training data used is controlled. 
We test the thief models trained on different amounts of data, either 100k, 400k, or the full available 8.7M training samples. 

The obvious observation to be made is that as the amount of training data is increased, the thief models better approximate the embedding models that they steal from. 
This is evidenced by the generally consistent improvement in retrieval scores across DL19 and DL20 shown in Table~\ref{tab:starve}.

The table also shows that the Cohere thieves score stronger than the OpenAI thieves. This is examined further in the following subsection.

\subsection{Cohere vs OpenAI}
\label{comparisoncohereopenai}
We observe that stealing from Cohere's model is easier than stealing from OpenAI's model. 
Examining in-domain retrieval effectiveness scores in Table~\ref{tab:starve}, the Cohere thief model trained on 100k training samples achieves higher scores than even the OpenAI thief model trained on 400k training samples.
While the Cohere thief trained on the full available training samples even scores stronger than the Cohere API model, the OpenAI thief lags behind its respective model. 
We suspect that the Cohere model is easier to steal from because it is initialized with a variant of a BERT model, possibly a BERT large variant as we have argued in Section~\ref{API}, and our thief models are initialized with BERT models as well.
We discuss the implications of this in Section~\ref{Defense}.

 \begin{table}[t!]
\centering
\resizebox{\columnwidth}{!}{ \begin{tabular}{l|c|cc}
\toprule
\toprule
 & \multicolumn{1}{c|}{\textbf{Cosine Loss}} & \multicolumn{2}{c}{\textbf{Contrastive Loss}} \\
   &  & \boldmath{$\tau=0.01$} & \boldmath{$\tau=0.05$} \\

\midrule
DL19        & 67.2 & 30.3 & 59.0 \\
DL20        & 67.7 & 31.6 & 61.0 \\
TREC-COVID  & 65.0 & 24.8 & 61.1 \\
NFCorpus    & 33.1 & 12.4 & 29.4 \\
SCIDOCS     & 14.3 & 4.8 & 12.3 \\
SciFact     & 65.5 & 23.1 & 60.0 \\
\midrule
\textbf{Average} & \textbf{52.1} & 21.2 & 47.1 \\
\bottomrule
\bottomrule
\end{tabular}
}
\caption{nDCG@10 scores comparing training with a cosine loss and a contrastive loss on TREC-DL and BEIR datasets for BERT base models distilled from Cohere's model with 100k training samples.
}
\label{tab:contrast}
\end{table}

\subsection{Contrastive Loss Ablation}
 \label{ContrastiveLoss}
Table~\ref{tab:contrast} compares training with a cosine distance loss and training with a contrastive loss, presented in Section~\ref{ContrastiveLossFunction}, as proposed in \citet{sha2023can}. Importantly, the work does not specify the temperature: $\tau$ value used for training. We test $\tau$ values of $0.01$ and $0.05$ as they are often used for the training of embedding models for retrieval \cite{ni2021large, gte, izacard2021contriever}. We find that training with a cosine loss achieves the strongest results. While a careful tuning of $\tau$ may see improved results, it is not clear whether the same $\tau$ value could be used across training settings regardless of the passages in the training data. It may be that the contrastive loss is not suitable when distilling using similar texts, where contrasting text representations could be counter-productive. Training with a cosine distance loss is a simple approach that works well without having to tune an additional parameter.

\begin{table}[t]
\centering
\resizebox{\columnwidth}{!}{
\begin{tabular}{l|cc|cc}
\toprule
\toprule
& \multicolumn{2}{c|}{\textbf{Cohere}} & \multicolumn{2}{c}{\textbf{OpenAI}} \\ 
& \multicolumn{1}{c}{\textbf{Final}} & \multicolumn{1}{c|}{\textbf{Bottleneck}} & \multicolumn{1}{c}{\textbf{Final}} & \multicolumn{1}{c}{\textbf{Bottleneck}} \\ 
\midrule
DL19        & 70.6 & 70.5 & 68.3 & 69.6 \\
DL20        & 72.2 & 70.7 & 68.9 & 68.8 \\
TREC-COVID  & 70.6 & 69.3 & 64.9 & 65.5 \\
NFCorpus    & 35.1 & 35.7 & 36.9 & 37.3 \\
SCIDOCS     & 16.7 & 17.1 & 15.3 & 15.6 \\
SciFact     & 67.1 & 68.2 & 65.7 & 65.7 \\
\midrule 
\textbf{Average}  & 55.4 & 55.3 & 53.3 & 53.8 \\
\bottomrule
\bottomrule
\end{tabular}
}
\vspace{0.2cm}
\caption{nDCG@10 scores for the BERT base thief models, comparing two settings: {Final Output Embedding} (1024-dimensional embedding for Cohere or 3072-dimensional embedding for OpenAI after the final layer) and {Bottleneck Embedding} (768-dimensional embedding before the final layer). Thief models encode both queries and passages.
}
\label{tab:bottleneck}
\end{table}

\subsection{Bottleneck Representations}
\label{BottleneckRepresentations}

As mentioned in Section~\ref{Architecture}, BERT base produces embeddings of dimension 768 and BERT large produces embeddings of dimension 1024.
This means that to distill from Cohere's and OpenAI's models, some transformation is needed to match the dimensions of the Cohere model at 1024 dimensions and the OpenAI model at up to 3072 dimensions, except for when distilling from Cohere's model to BERT large. 
For transformations, we use a simple linear mapping. As such, the intermediate representation of fewer dimensions can be normalized and used as the embedding for retrieval. 

Table~\ref{tab:bottleneck} demonstrates that these shorter embeddings yield retrieval effectiveness comparable to that of the thief models' full-length embeddings after the final linear mapping.
For this reason, we use the shorter embeddings to assess retrieval effectiveness when possible. 
This finding implies that the distillation method presented in this work can effectively transfer retrieval effectiveness from models with higher-dimensional embeddings to models with lower-dimensional embeddings.

\subsection{Distilling from Both Models at Once}

We explore whether an embedding model can benefit from distilling from both Cohere and OpenAI embeddings.
The aim is to train a student model that captures and benefits from the relevant information in both embeddings. 
To study this, we concatenate the vectors from Cohere's and OpenAI's embedding models and attempt to distill these concatenated embeddings into a BERT large thief model.
As detailed in Table~\ref{tab:joint}, the results are promising.
For the MSMARCO and BEIR tasks examined, using the concatenated embeddings for retrieval achieves higher nDCG@10 and R@100 scores than using either set of embeddings alone.
Then, when we distill using the concatenated embeddings into our thief model, the thief model achieves a higher average nDCG@10 score than when distilling from either API embedding model alone.
Further, the thief model often scores stronger than either API model alone. 
 
This means that taking embeddings from leading embedding models, concatenating them, and then distilling these concatenated embeddings into a student model is a promising method to train embedding models. 
This approach possibly allows for the integration of diverse model strengths in creating more robust embedding models. 

\begin{table}[t]
\centering
\resizebox{\columnwidth}{!}{
\begin{tabular}{l|cc|cc|cc}
\toprule
\toprule
 & \multicolumn{2}{c|}{\textbf{Cohere}} & \multicolumn{2}{c|}{\textbf{OpenAI}} & \multicolumn{2}{c}{\textbf{Concatenate}} \\
 & \multicolumn{1}{c}{\textbf{API}} & \multicolumn{1}{c|}{\textbf{Thief}} & \multicolumn{1}{c}{\textbf{API}} & \multicolumn{1}{c|}{\textbf{Thief}} & \multicolumn{1}{c}{\textbf{API}} & \multicolumn{1}{c}{\textbf{Thief}} \\

\midrule
DL19        & 69.6 & 70.6 & 71.7 & 68.3 & 75.0 & 73.2 \\
DL20        & 72.5 & 72.9 & 71.6 & 69.3 & 74.5 & 72.4 \\
TREC-COVID  & 81.8 & 73.5 & 76.9 & 68.7 & 83.1 & 70.0 \\
NFCorpus    & 38.6 & 36.6 & 41.8 & 38.9 & 42.4 & 39.8 \\
SCIDOCS     & 20.3 & 17.5 & 22.8 & 17.3 & 22.8 & 18.0 \\
SciFact     & 71.8 & 69.3 & 76.1 & 67.5 & 77.6 & 72.8 \\
\midrule
\textbf{Average}  & 59.1 & 56.7 & 60.2 & 55.0 & \textbf{62.6} & \textbf{57.7} \\
\bottomrule
\bottomrule
\end{tabular}
}
\caption{nDCG@10 scores on TREC-DL and BEIR datasets, using a {Concatenate} setting where Cohere and OpenAI embeddings are combined for retrieval and distilled to a BERT-large thief model. The thief model is used to encode both queries and passages.
}
\label{tab:joint}
\end{table}

\subsection{Cost to Steal}

We note that at the time of this study, a VM with a 48GB RTX 6000 Ada GPU can be rented for \$0.74/hr.\footnote{\url{https://www.runpod.io/gpu/6000-ada}} This puts the cost of training the BERT large models for 268 hours at roughly \$198 and the cost of training the BERT base models for 104 hours at roughly \$77. 
We estimate the cost of generating the embeddings in our training set and dev set at roughly \$88 using OpenAI’s text-embedding-3-large model and roughly \$68 using Cohere’s embed-english-v3.0 model.
This means that our model-stealing can be done with a modest cost of under \$300 before applicable taxes.

\section{Defense}
\label{Defense}

Our results indicate that Cohere's embedding model is easier to distill into a BERT thief model than OpenAI’s, likely because Cohere's model is based on a BERT variant, as noted in Section~\ref{API}.

It may be easier for an adversary to steal an embedding model if they can correctly guess the model backbone of the embedding model and initialize their thief model with the same backbone. 
For this reason, we recommend that commercial embedding models be initialized with some model such that it would be difficult or impossible for an attacker to guess the model and initialize their model with the same architecture and weights. 

We also recommend that companies take care to not expose key model details unnecessarily. 
As we have explained, we can presume that Cohere uses a BERT variant for initializing their embedding model because they make their tokenizer publicly available through HuggingFace, where we can see that the tokenizer is a \texttt{BertTokenizer}.

Moreover, Cohere provides embeddings for the MSMARCO v1 corpus and BEIR datasets,\footnote{\url{https://huggingface.co/datasets/Cohere/beir-embed-english-v3}} which, while valuable to the research community, also makes it easier for an adversary to steal their model by providing training data for distillation free of cost.

\section{Conclusion}

In this work, we were able to successfully distill commercial embedding models from behind their APIs to BERT thief models. 
These thief models demonstrate strong effectiveness across MSMARCO and BEIR benchmarks, generalizing well to out-of-domain tasks despite training exclusively on MSMARCO v1 passage ranking data.
We find that these stolen embedding models can be used to embed both queries and passages for retrieval with strong effectiveness. 

We also find success in experimenting with training stronger embedding models by distilling from multiple strong embedding model teachers into a single student model. 
Additionally, we show and empirically verify a simple way to distill from embedding models with high-dimension embeddings to models that produce embeddings with fewer dimensions. 

Our findings expose the susceptibility of commercial embedding models to theft. 
We highlight the need for greater consideration for designing commercial embedding models that are less prone to being stolen by attackers. 
Through our results, we arrive at the recommendation to initialize commercial embedding models with less predictable backbones, unlike BERT, which may be more vulnerable to theft. 
Employing unique or customized model backbones can potentially increase the difficulty for attackers attempting to steal the model.

Looking forward, further research may investigate improving the effectiveness of the thief models by employing more extensive and diverse text training data for distillation. 
Future research must also further investigate defending against model theft. Additionally, there is an opportunity to investigate distillation from multiple strong teacher embedding models to develop more robust and efficient student models. These explorations could provide valuable insights into the competitiveness of commercial embedding models in a rapidly evolving space.

\section*{Acknowledgments}
We thank N.\ Asokan, Ian Goldberg, and Florian Kerschbaum for their guidance on the ethical considerations of model stealing. 
This research was supported in part by the Natural Sciences and Engineering Research Council (NSERC) of Canada and Microsoft via the Accelerating Foundation Models Research program.
We also thank Compute Canada for providing computing resources. 

\bibliography{custom}

\clearpage
\newpage

\appendix

\section{Limitations}

\paragraph{Student Model Initialization.} 
We only study using BERT base and BERT large to initialize our student models. We do this to run our experiments quickly and with ease. However, further experiments may find that student models initialized using a more recent LLM may be able to better replicate these embedding models behind APIs, especially if the models behind the APIs are themselves initialized with LLMs. These student models may be able to attain stronger retrieval effectiveness scores with less training data needed.

\paragraph{Training Data.}
Another limitation is that we train only using queries and passages from the MSMARCO v1 passage ranking task. There are some benefits and drawbacks to this. Notably, passages from the MSMARCO v1 passage corpora tend to be relatively short, generally consisting only of a couple of lines. Training with shorter passages allows for faster training and cheap encoding, since the APIs charge for encoding per token. However, this is at the potential cost of worse effectiveness when encoding longer texts. While we note that our student models are able to generalize well to diverse BEIR tasks, some with longer texts, they may still suffer when being used to encode much longer texts. Regardless, training with more text--embedding pairs and with more diverse text--embedding pairs can mitigate these concerns.

\paragraph{Exploring Defenses Against Model Theft.} 
We recognize the need to propose and test more effective defenses against model theft. However, we leave this for future work as this is a difficult problem. With stealing sentence encoders, \citet{dziedzic2023sentence} studied watermarking~\cite{watermarking}.
With stealing image encoders, both \citet{sha2023can} and \citet{ stolenencoder} also explored watermarking and perturbation-based
defense. However, watermarking only serves to
identify stolen models~\cite{watermarking} and both
works found that perturbation-based defense was not
effective for defending against stealing attacks for
encoder models because the perturbations could not sufficiently hurt the effectiveness of the attack
while maintaining the effectiveness of the victim
model~\cite{sha2023can, stolenencoder}.

\section{Ethical Considerations}

We informed Cohere and OpenAI of our success in distilling their embedding models from their APIs over one month before publicly posting our work.
We acknowledge the legitimate interest of companies in profiting from their proprietary embedding models, which require significant effort and investment to develop.
Bad actors may try to steal commercial embedding models to profit from them. Further, adversaries may try to design adversarial attacks using the stolen models with the hope that the attacks transfer to the original models.
However, we believe it is crucial to highlight that these models can be accurately and cost-efficiently stolen.
This is necessary to begin to study defense considerations against model theft such as what we have discussed in our paper. 

To mitigate potential misuse, we do not make our models or training code publicly available.
However, we are willing to provide access to researchers upon request, ensuring that the research community can benefit from our findings while minimizing the risk of unethical application.

\end{document}